# Crowd simulation for crisis management: the outcomes of the last decade

George Sidiropoulos[1], Chairi Kiourt[2], Lefteris Moussiades[1]

[1]Department of Computer Science, School of Science, International Hellenic University, Kavala, Greece
{georsidi, lmous}@teiemt.gr
[2]Athena-Research and Innovation Center in Information, Communication and Knowledge Technologies, Xanthi, Greece
chairiq@athenarc.gr

**Abstract**: The last few decades, crowd simulation for crisis management is highlighted as an important topic of interest for many scientific fields. As the continues evolution of computational resources increases, along with the capabilities of Artificial Intelligence, the demand for better and more realistic simulation has become more attractive and popular to scientists. Along those years, there have been published hundreds of research articles and have been created numerous different systems that aim to simulate crowd behaviors, crisis cases and emergency evacuation scenarios. For better outcomes, recent research has focused on the separation of the problem of crisis management, to multiple research sub-fields (categories), such as the navigation of the simulated pedestrians, their psychology, the group dynamics etc. There have been extended research works suggesting new methods and techniques for those categories of problems. In this paper, we propose three main research categories, each one consist of several sub-categories, relying on crowd simulation for crisis management aspects and we present the outcomes of the last decade, focusing mostly on works exploiting multi-agent technologies. We analyze a number of technologies, methodologies, techniques, tools and systems introduced throughout the last years. A comparative review and discussion of the proposed categories is presented towards the identification of the most efficient aspects of the proposed categories. A general framework, towards the future crowd simulation for crisis management is presented based on the most efficient to yield the most realistic outcomes of the last decades. The paper is concluded with some highlights and open questions for future directions.

**Keywords**: Crowd simulation, crisis management, multi-agent systems, machine learning

## 1 Introduction

The advent of the web, computational resources and the Artificial Intelligence (AI), became an initial motivation for researchers in the area of Crowd Simulation (CS). In general, the CS research field is growing more and more in the last decade. As a consequence, there have been many techniques and methods proposed for different research simulation sub-fields. Firstly, there have been many methods for the simulation of crowd motion/interactions in different situations. For example, there has been a focus on crowd behavior simulation ([1]), emotion contagion ([2], [3]), collision avoidance ([4], [5]) and other. Moreover, due to the fact that the performance and broader use of Machine Learning (ML) and Deep Learning (DL) techniques has increased, there have been many methodologies that take advantage of those. Reinforcement Learning (RL) is another important subfield of ML with many research outcomes in the area of CS [6], [7] and [8].

The interest in the field of pedestrian crowd management was present as early as 1958 [9] and has been continually studied for all those years [10]–[12]. The main goal of these studies was to develop a level-of-service concept, design elements of pedestrian facilities or planning guidelines [13]. With the emergence and increasing exploitation of crowd simulation systems the goals have remained the same, but the demand has increased and continue to increase day by day. The construction of large scale, or even small-scale, buildings requires extensive and correct planning, that aims for not only style and appearance, but also for functional requirements and visitor behavior scenarios [14]. One important behavior scenario is a crisis scenario (e.g. evacuation due to fire), which aims to improve the procedure for risk assessments, emergency plans and the evacuation itself. Crisis management preparation procedures used today include fire drills and “mock evacuations”, but those generally fail to accurately prepare us for these kind of scenarios and are very often ignored [15]. Thus, we cannot design accurate policies based on the results from those preparations. For this reason, simulations can provide an additional method of evaluating security policies that take into account the impact of different environmental, emotional and informational conditions [16].

The past years, there has been an increasing interest towards the research field of Crowd Simulation for Crisis Management CSCM) related algorithms. Crowd simulation is the process of simulating how a number of entities (commonly large) move inside a virtual scene with a specific setting [17]. The type of setting can vary, from film production and military simulation to urban planning, which require high realism concerning the movement patterns and grouping of the entities. Crisis simulations, or crisis management systems, are systems which include not only entities with more roles and responsibilities but also more techniques and algorithms that are responsible for the physical and also psychological simulation of those entities.

For the purpose of simulating multiple individual entities, Multi-Agent Systems (MAS) are considered to be the most suitable architectures/systems [18]. MAS are consisted of agents (entities to be simulated) and their environment (the setting in which they exist) that solve specific tasks [19]. By exploiting their knowledge, agents interact with other agents in the environment, or with the environment itself. Based on their interactions and their perception agents perform actions to achieve their goal. Their structure makes them befitting for crowd and crisis simulation research.

The motivation of this research is to present the development and outcomes of the field of CSCM, the new technologies that have been presented and proposed, along with the new capabilities that have emerged through the research that has been conducted. Moreover, this field continues to grow every day as it is very active, with new systems and technologies being released as the time goes by, with researchers having to be up to date about all the new advancements.

The main contribution of this paper is to provide a comprehensive review of the literature of the last decade. In this review, we divide the literature into different categories (research sub-fields of CSCM), giving the reader a structured overview of the methodologies and available tools for crowd and crisis simulation. Furthermore, the goal of this work is to propose a general framework for the development of a multi-agent system for CSCM, taking into account all the strong points and drawbacks derived from the reviewed literature. The framework will enable the user to exploit cutting-edge simulation technologies for the management of crisis situations, including the entities that have to be parameterized.

The remainder of the paper is organized as follows: Section 2 shows the methodology that was followed to review the literature, giving an idea of how the papers presented in this review were selected. Section 3 outlines detailed thoughts about crowd simulation and crisis management and other related works that have been done in the literature. Following, Section 4 presents the analysis of the methodologies, systems, tools and techniques that have been presented and proposed. In Section 5 a comparison of those works is presented, in Section 6 the key elements of a general framework are presented, followed by Section 6 that concludes the review, along with the future direction of the literature and some key points that have to be taken into account.

## 2 Methodology of Literature Review

The first step of the literature review was to gather a large sample of published works concerning multi-agent crowd or crisis simulation frameworks, as well as proposed techniques that aimed to improve the performance, realism or in general improve the quality of the simulations. Aiming at these types of works, a more comprehensive view of the literature can be derived, including its trends and concerns, collecting a large enough sample that can represent it.

The main source of the literature was gathered from SCOPUS[1] in conjunction with Google Scholar[2]. Apart from these databases, other sources were also searched using the references of the relevant articles, giving the ability to also track relevant conference proceedings. As mentioned before, the literature review focused on the works published in the last decade (about 2009 to 2020), but also included a few earlier works, without having a limit on the number of gathered publications. Regarding the keywords, there was a number of keyword queries used to restrict out-of-scope publications, such as:

- "**crisis simulation AND (agent OR artificial OR multi-agent OR evacuation)**",
- "**crowd simulation AND (agent OR artificial OR multi-agent OR evacuation)**" and
- "**(crowd simulation OR crisis simulation) AND reinforcement learning**".

Thus, the total number of the publications obtained was 92. The distribution of the gathered number of works per year is shown in Figure 1 (a). The inclusion criteria of the works collected was, firstly, the total number of citations to-date the presented work had and then the presentation of the results and methodology. It should be noted that, due to the fact that the review is focused on the last decade, the citation criteria did not apply to works published on the last 3 years (2018, 2019 and 2020), as they are new and the low citations number is natural.

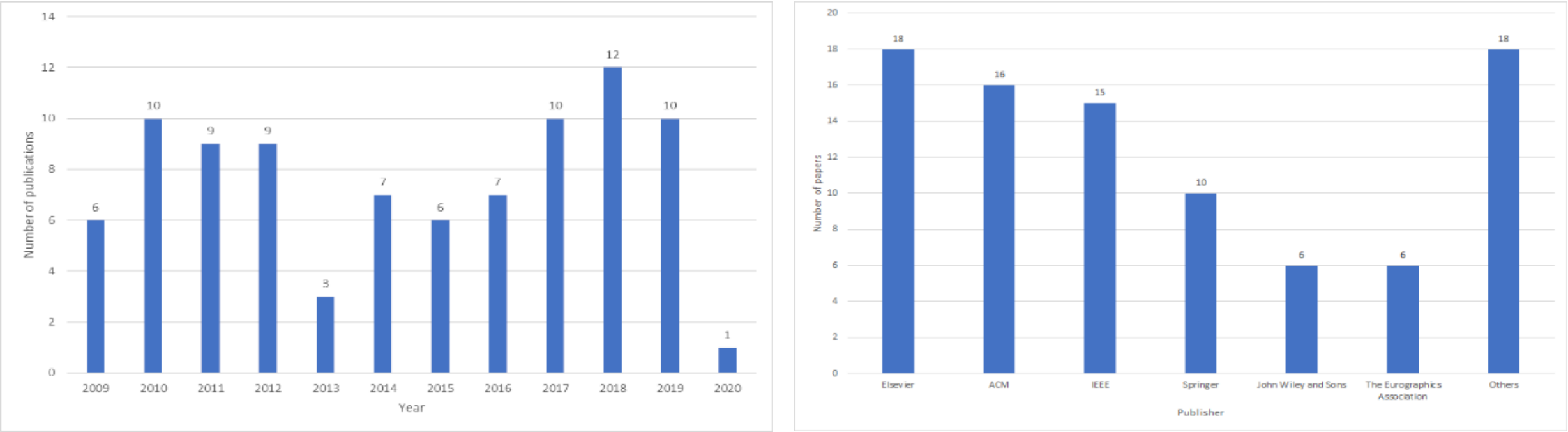

(a) (b)

Figure 1 (a) Number of publications gathered for each year. (b) Publishers and number of papers collected from each

Figure 1 (b) depicts the distribution of the publications based on the publishers. Most of the papers were published in Elsevier (where most of the search was done), followed by ACM Digital Library with 16 and then IEEE with 15. The "Others" bar in the graph of the Figure 2 (b), include papers from thesis or papers with no official proceedings of publishers.

# 3 Background and Related Works

There has always been a deep interest in crowd simulations, as someone can derive a deeper understanding of the crowd's movement, motion and behavior though it. Crowd behavior is complicated and depends on the decisions of each individual, or of a group as a whole, while also taking into account their surroundings (obstacles, other individuals or environment) and their goal. Taking those into account, along with the fact that crowd simulation requires cognitive modeling and rendering, the simulation becomes quite difficult. For this reason, crowd simulation has been studied extensively even in other fields, such as social sciences, traffic engineering, architecture etc. As early as 1976, Tsuchiya et al. simulated the results of three transportation systems in a certain city, providing computational results with and without those transportations [20]. Those simulations were done by using mathematical models to describe parts such as zoning of urban area, search of trip, trip allocation etc. Similar mathematical models were still being used years later. For example, [21] simulated the behavior and movement of the crowd during the Tawaf[3] using mathematical relations for the radial and lateral movement and stoning process. In 1999, Jiang simulated pedestrian movements in urban environments and found that the morphological structure of the environment has striking impacts to the movement [22].

[1] https://www.elsevier.com/solutions/scopus, The largest database of peer-reviewed literature - Scopus | Elsevier Solutions

[2] https://scholar.google.com/, Google Scholar

The behaviors simulated in a crowd simulation system can be viewed and analyzed at *microscopic* and *macroscopic* levels. At the microscopic level, crowd simulation is dealing with the behavior of, or generally focuses on, each entity individually. On the other hand, at the macroscopic level, the simulation is focused on the behavior of the crowd or groups of crowds as a whole.

Crowd simulation has also been extensively used in crisis settings, where the crowd and its evacuation process are simulated in different kinds of environments and scales of crisis. While the frequency of large-scale crisis situations is low and studying the crowd's behavior is difficult, due to the fact that it requires either exposing people to dangerous situations or recreating drills (which are not taken into account seriously by people), the use of simulation tools becomes essential. The importance of the simulation in those scenarios becomes more challenging when the psychological element of the crowd is taken into account. In 1977, Hirai and Tarui proved that people under panic start to follow groups of people, fail to use the evacuation means effectively and display herding behaviors [23]. Moreover, Cardon and Durand presented a dynamic model for crisis management that took into account the mental representations of the actors, allowing the simulation of intentions and judgements when the actors exchanged information about the situation [24].

A very common approach to the modelling of the simulations is the Agent-Based Modelling (ABM), in which each entity (for example an individual pedestrian) is represented by an agent. An agent is an autonomous entity, which has a specific set of goals and takes actions based on interactions with the environment and its elements towards the achievement of its goals with the highest profitability [25]. Due to the autonomous nature of this modelling approach MAS provide a more natural simulation in comparison to other approaches that use mathematical models to predict the behavior of the crowd. MAS are also extensively used for this purpose and there are numerous frameworks and tools released throughout the years, such as NetLogo [26], MATSim [27], MASSIS [28], DrillSim [29] and the GAMA platform [30]. A very early simulation tool that used multi-agent modelling was presented by A. Drogoul et al. [31] and aimed to simulate different species in an environment, demonstrating its use on simulating an ant colony.

Though the years the focus of the proposed methodologies has remained almost the same. In 2001, Goldenstein et al. focused on the crowd's movement [32], specifically 3D steering and flocks among obstacles. Moulin et al. built a geosimulation system for the simulation of a several thousands of agents [33]. Law et al. focused on the human decision-making and interaction by building a framework to study those [34]. [35] presented a system for simulating individually hundreds of people, focusing on the interactions between them. [36] also focused on the interaction between agents by presenting a multi-agent framework. [37] focused on the generation of feasible paths for crowds by also maintaining a specific shape and, similarly, [38] focused on real time path planning and navigation of multiple agents in dynamic environments.

Figure 3 shows a word cloud created from the keywords of all the papers included in this review. From the figure we can highlight that the "simulation", "crowd simulation" and "evacuation" keywords are largely used, naturally, along with the "agent", or "multi agent" which means that the multi-agent-based methods are quite highly exploited in CSCM cases.

---

[3] The Tawaf is one of the Islamic rituals of pilgrimage in Mecca, The Hejaz, Saudi Arabia, performed during the Umrah and Hajj. During the Tawaf, Muslims go around the Kaaba seven times, in a counterclockwise direction; the first three circuits at a hurried pace on the outer part of the crowd, followed by four times closer to the Kaaba at a leisurely pace. (https://en.wikipedia.org/wiki/Tawaf)

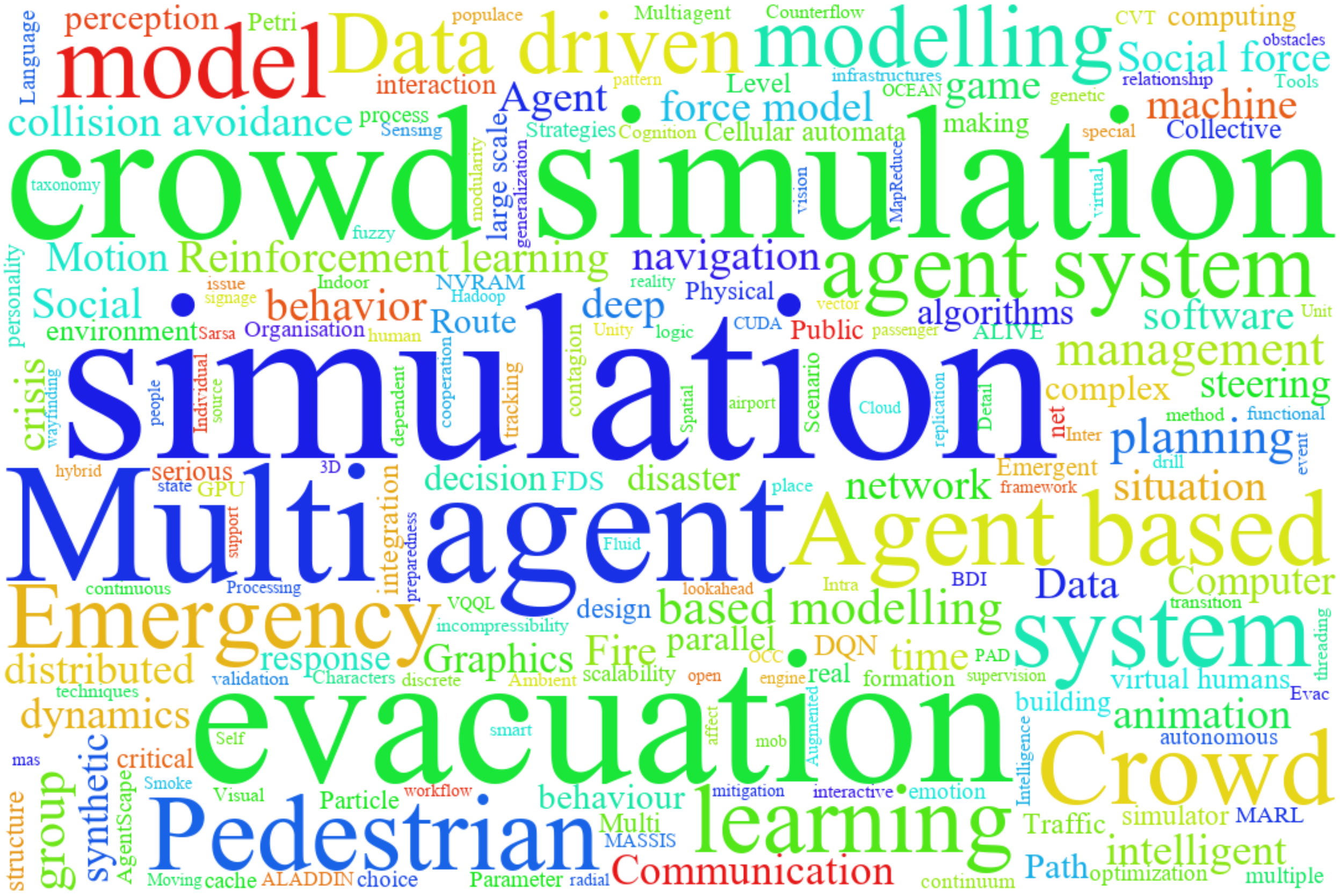

Figure 2 Word cloud generated using the keywords of the reviewed literature.

In the literature there is a number of similar review and survey papers focusing on crowd simulation, crisis simulation or crisis management. Legget reviewed the area of real-time crowd simulation, describing the three major approaches to this problem [39]; Fluid-based [40], Cellular Automata (CA) [41] and Particle-based [42]. In addition to that, they presented the CrowdSim [43], a simple implementation of some techniques. Heath et al. did a survey on agent-based modeling (ABM) from 1998 to 2008, collecting data from 279 articles, to establish the practices of ABM in terms of simulation software: purpose of simulation, acceptable validation criteria, validation techniques and other data [44]. They also recommended some improvements towards the future advance of ABM systems, as it was an immature method then. Duives et al., reviewed existent pedestrian simulation models of the last decade, to ascertain whether those models could be used for simulating high density crowds, arguing that any crowd simulation model should be able to simulate most of the phenomena indicated [45]. Bañgate et al. reviewed the evacuation behavior, theories on social attachment, crises mobility and agent-based models [46]. They showed the trends in behavior modelling for crises, the theories related to social attachment and introduced 12 agent-based models that implemented social elements. Recently, Y. Li et al. provided an overview of the CA evacuation models and presenting the main challenges that exist for these models along with their advantages and disadvantages [47].

Compared to the aforementioned reviews and surveys, we have done a more general review of the literature, focusing on aspects and parts of CSCM systems (methods and techniques, simulation models, systems and tools) related to multi-agent methodologies/techniques. Moreover, focusing on the last decade, along with the different aspects, we provide a broader view of the trends and outcomes, helping those who are interested field and want a clear view of the field of CSCM. Lastly, the comparative and categorical analysis of the publications focus on assisting scientists towards the development of novel CSCM systems, algorithms and theories based on a general multi-agent framework.

## 4 Methods, Techniques, Models, Systems and Tools

In this section we analyze all the methods, models, systems and tools of the last decade in the field of CSCM, separated into three main categories: *Systems and Tools (ST), Simulation Models (SM) and Methods and Techniques (MT).* Each main category consists of some sub-categories. The first category (ST) is divided into what settings the systems or tools aim to simulate, that is: Crowd simulation and Crisis simulation, while the third category (SM) is divided into the specific simulation parts of a CSCM model: Navigation, Agent behavior, Emotional aspects and Group dynamics. Figure 3 depicts the entire structure of the categorization of CSCM problems/research subfields.

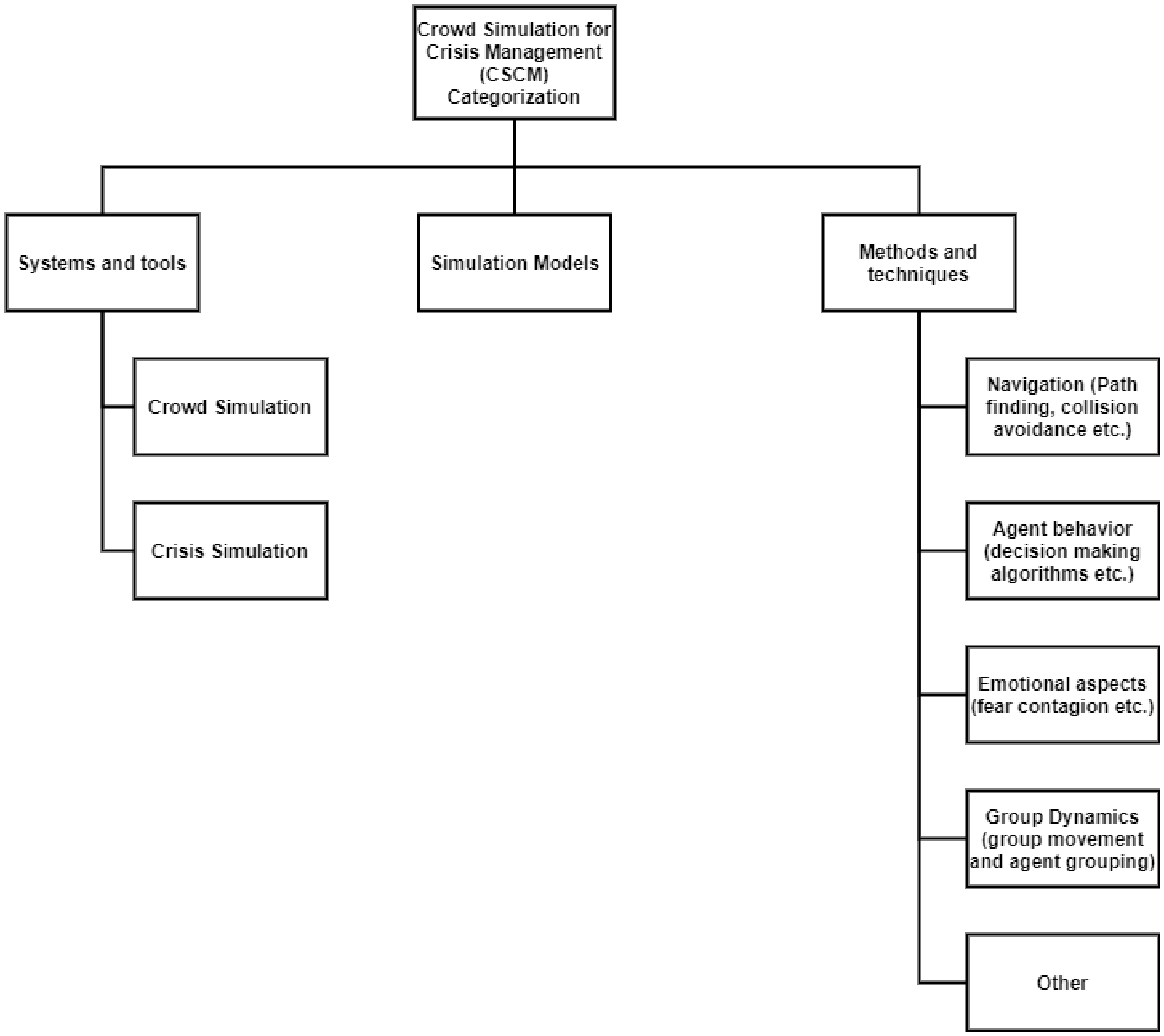

Figure 3 Organization chart showing the categories and sub-categories of the publications included in this review.

The reasoning behind the way we divided the works is due to the fact that there are many proposed methods and techniques for faster and efficient computation of different parts of a simulation procedure. For example, there are many algorithms for the computation of the closest path to a target or for the grouping of the crowd. Furthermore, there have been a number of proposed models that aim to simulate crowds or crisis situations combining different, newly proposed or not, physical, mathematical and other models, but use a pre-existing simulation tool or framework, or one that was developed only for its evaluation. Lastly, the last category presents some of the most popular different simulation systems and tools that were developed and proposed for CSCM.

## 4.1 Systems and Tools

In this section, we present chronologically the crowd and crisis simulation systems and tools that have been published. Systems and tools are those that can be used to fully simulate a scenario, whether it can simulate a specific scenario only, e.g. Tawaf, or a more generic scenario, like a fire building evacuation.

### *4.1.1 Crowd simulation*

A novel multi-layered flocking system with the ability to let agents of vastly different sizes move underneath another, when there was enough space, was proposed in 2010 by Van Den Hurk & Watson [48]. The system is based on the original Reynold's flocking model [49], in addition of a series of layers to represent the simulation space. The proposed system allowed more complex navigation simulations, the traversable space could be partitioned and the characters behavior was highly parallelized. Despite those, it inherited the problems existing in Reynold's method, its efficiency was based on an assumption (a smaller object navigates around a larger) and the agents required information about other layers which were not naturally handled by the system.

A system that focused on simulating the movement of individuals in large scale crowds performing the Tawaf was introduced in 2011 by Sean Curtis et al [50]. The proposed approach used a finite state machine and an agent-based algorithm, along with the Reciprocal Velocity Obstacles (RVO) algorithm [51] for obstacle avoidance. The simulation showed that the system could run in real-time for a high number of agents and the results matched those observed in real life. On the other hand, no groups are simulated, there was an unknown impact of heterogeneity and the agent's speeds needed to be validated.

Unity game engine[4] is one very popular simulation development platform that has been exploited for several CSCM studies. In 2014, [52] focused on first-persona perception and signs, taking dynamically changing occlusions into account. They also used an online recalculation of sign visibility from the perspective of each agent. The implementation was done using Unity, while also making it possible for participants to be tested in the same virtual airport terminal, with the combination of a head-mounted display "Oculus Rift". This system accurately simulated 3D perception and interpretation of signage information. On the other hand, the agents did not line up to pass checkpoints or ask airport personnel for help and got distracted by advertisements and shops.

CA based approaches are some of the earliest attempts trying to simulate CSCM cases, but still very popular. In 2015, Suzumura et al. aimed to simulate billions of agents at a microscopic level and proposed a complex agent-based CA architecture [53]. The simulations were done using the X10 programming language and used the ScaleGraph library [54] to enhance the program's performance. They simulated a running traffic flow with a billion agents (using the TSUBAME supercomputer with 1536 CPU cores) but the simulations were not very realistic. Similarly, Yu et al. proposed a method that used CA and multi-agent models [55], which were mapped onto the MapReduce[5] programming model [56]. In addition to those, they developed a framework that utilizes Apache Hadoop[6] [57], in order to simulate large crowds in a generalized environment model. Compared to the serialized simulations, the framework run much faster with reduced memory working set size, with an increase in total memory usage. Additionally, the simulation process was in need of better control mechanisms and an improved environment model.

Menge is an agent-based, cross-platform, extensible, modular framework for simulating pedestrian movement in a crowd [58]. Sean Curtis et al. simulated the crowd movements by subtracting the problem into subproblems, while also simulating human response to stress, crowd formation, density-dependent behavior and formation stress. A main drawback, was that their simulations had a fixed population and that the framework lacked scalability. Focusing on a different aspect, [59] aimed to create a system for the specification of different types of audience and mob crowds, associating psychological components with individual agents comprising a crowd. They used the OCEAN (Openness Conscientiousness Extroversion Agreeableness Neuroticism) [60] and OCC (Ortony Clore Collins) [61] models for the personality and emotion synthesis respectively, a generalized emotion contagion method and the Pleasure-Arousal-Dominance model [62] to determine the current emotional state of an agent. Lastly, they used the Unity Game Engine for the simulations. In 2017, Malinowski et al. introduced a custom modular, parallel, agent-based for large-scale crowd simulation environment [63]. The system was implemented in C, using the Message Passing Interface[7] (MPI) [64] for parallel environment simulation and visualization. From the simulation results, it was shown that the architecture was able to perform simulations of thousands of agents on an area of districts or even cities with many future promising outcomes.

[4] A game engine developed by Unity Technologies (https://unity.com)

[5] A programming model and an associated implementation for processing and generating big data sets with a parallel, distributed algorithm on a cluster.

[6] A collection of open-source software utilities that use a network of many computers to solve problems involving massive amounts of data and computation.

Another powerful and very popular game engine exploited for simulation applications is Unreal Engine. In 2018, Simonov et al. proposed a system for building composite behavior structures for large number of agents [14]. This system was consisted of a decision-making system, implemented in Unreal Engine 4[8] Blueprints, the path finding system used the Menge simulation with plugins and the system also included animation support, dynamic models, a visualization module and utility-based strategic level algorithms. The scenario simulated was of an Olympic Park train station in Sochi, during Winter Olympics 2014, simulating over 1700 agents.

In 2019, Malinowski and Czarnul presented a novel multi-agent based architecture for the development of a parallel and modular agent-based environment, along with its architecture and main components, it incorporated an Non-Volatile Random Access Memory (NVRAM) device while also using MPI I/O (Input/Output) calls to communicate with an application state [65]. The proposed systems could simulate an environment of more than 60.000 agents, but the high number of I/O requests increased the execution time and there was a lack of fault-tolerance mechanisms. Lastly, another very popular platform that adapts to large-scale and complex models is the GAMA platform [30]. Currently in its 1.8 version, GAMA platform has its own agent-oriented modeling language called Gama Modeling Language (GAML) that follows the object-oriented paradigm. Additionally, the models include spatial components used to represent their 3D representation in the environment. Furthermore, another key feature of the platform is the agent's architecture, based on the Belief Desire Intention (BDI) paradigm [66], that proposes a straightforward formalization of the human reasoning through intuitive concepts. It also supports multi-threaded simulations and running multiple simulations at the same time.

#### *4.1.2 Crisis simulation*

Temporal-Difference Learning (T-DL) [67] algorithms, a core branch of Reinforcement Learning, are some of the most popular approaches exploited in CSCM applications. In 2009, [68] used the T-DL to develop a basic large building evacuation plans for a non-stationary environment simulating fire, composed of heterogeneous population (multi-agent system). On one hand, the created program was versatile, having a range of simulations it could run. On the other hand, the way the building was designed lead to different routes to the exits having inaccurately the same distance, people did not respond immediately to an evacuation alert, there was a small variety of agents that had no physical space requirements and action-selection algorithms. Shi et al. presented an evacuation simulation system, called AIEva, which included a physical and mathematical model that aimed to simulate crowd evacuation in fire environments for large buildings [69]. It used agent simulation techniques and other artificial intelligence methods, adopting different behavior rules and changes to occupant status due to the fire's harmful effects. The system, though, lacked flexibility and validity with no satisfying results.

---

[7] A standardized and portable message-passing standard to function on a wide variety of parallel computing architectures.

[8] A game engine developed by Epic Games. (https://www.unrealengine.com)

The adaptation of an existing multi-agent transportation simulation framework to large-scale pedestrian evacuation simulation was presented by Lämmel et al. in 2010, [70]. The simulation framework that was used, was based on the MATSim framework [71] for transportation simulation. The presented microscopic simulation framework was implemented as a multi-agent simulation system, where each agent tried to optimize the evacuation plan in an iterative way. The simulations conducted gave plausible results and the performance was suited for large-scale scenarios, but the simulation assumed that all evacuations were done by foot (pedestrians), without vehicles. The same year, Dimakis et al. presented a distributed building evacuation plan which provided mechanisms for the interaction of the simulated entities [72]. The system could incorporate different types of entities along with their interactions and strategies but lacked regarding the user interface and also required performance improvements

ESCAPES, a multi-agent evacuation simulation system, was presented in 2011, which incorporated different agent types with emotional, informational and behavioral interactions [16]. The agent types were consisted of individual travelers, families, authority and security agents. The spreading of the information to the agents, was done with the *exit knowledge method* (no information when entering a room for the first time) *and event knowledge method* (delay of the information of an event to agents away from its location). The emotional interaction was consisted of an emotional contagion process, in which the emotion was spread by agents to agents (transferring high level of emotion) or agents to authority agents (reducing their fear). For the behavioral interaction they used Social Comparison Theory [73], for cases like when an agent wished to urgently exit an area without knowing the exit's location. They simulated a scenario of the Tom Bradley International Terminal at Los Angeles International Airport, presenting results of scenarios with increased number of authority entities and exits. A year later, Ribeiro et al. presented a serious game evacuation simulator, where the player has to evacuate the building in the shortest time possible [74]. They created realistic 3D models of the environment in Blender[9] that were loaded onto the Unity3D game engine. Although the behavior of the subjects that was taken into account was realistic, it could be taken as a drawback as the simulation was not automated. On the other hand, in 2012 a novel fire emergency simulation system was introduced, that exploited graph mining, social network analysis and other agent-based techniques, called EvaPlanner [75]. The simulation system identifies preferable exit locations for efficient evacuation, the most efficient location for evacuation signs and tool into account individual kinetics and social connections between people. The simulations showed that the system was robust. In 2012, Evakuierungsassistent (translated as Evacuation Assistant) was presented as a simulation environment for the evacuation of mass events, incorporating realistic methods for the real-time simulation [76]. The system was agent-based and exploited CA methods and Generalized Centrifugal Force Models [77], optimizing them to run in real-time.

A simulation system to study the crowd's behavior while evacuating of a soccer stadium was proposed by De Oliveira Carneiro et al. in 2013 [78]. The authors, exploited the use of 2D CA defined over multiple grids that represented different levels (state spaces) of simulated environment. The individuals made better use of space available by moving and interacting with a small set of rules. Also, the system had the ability to simulate environments with complex structures composed of multiple floors. An Agent-based Decision Support System was introduced in 2014 by Wagner and Agrawal, that simulated the evacuation of a crowd during a fire disaster, using a decision support system that exploited agent-based modeling [79]. The system allowed the user to setup an environment and included fire dynamics and pedestrian moving algorithms that were consisted of exit selection, movement towards a pathway and movement along a path toward the exit. This system was a proof of concept at the time. Also, it had a simple person's decision-making process, the fire dynamics of the environment could be improved automatically and the system had the ability to model multilevel venues but without people moving between them.

[9] A free and open-source 3D computer graphics software toolset. (https://www.blender.org)

A crowd evacuation system, focusing on mass assemblies of pedestrians in Hajj rituals[10], was introduced by Mahmood et al. in 2017 [80]. The system also included an analysis framework that incorporated the simulation environment of AnyLogic Pedestrian Library[11],. Additionally, for the optimization they used Shortest Regional Distance and a Genetic Algorithm [81] . During the simulation they compared different crowd evacuation strategies and evaluated their performance.

Real world evacuation data were exploited for the training of a deep neural network that predicted the human behavior, depending on the surrounding situation. In 2018, Tkachuk et al. presented a program that solved practical problems connected with emergency evacuation from buildings using system simulation based on Artificial Neural Networks (ANNs) [82]. In 2019 Sharm et al., proposed the first fire evacuation environment based on the OpenAI gym[12] [83] [7]. Moreover, they proposed a new approach that entailed pretraining an agent based on a Deep Q-Network (DQN) algorithm [84] focusing on the discovery of the shortest path to the exit. The proposed approach had faster convergenecy and training time, was scalable and had higher stability compared to others. The same year, a new RL based data-driven crowd evacuation framework was presented by Yao et al., to enhance the visual realism of crowd simulation [85]. The system extracted dynamic characteristics from videos. Also, a K-Means based model with a hierarchical path planning method was used to group the crowd and to merge the individual's trajectories. The group trajectories were computationally more efficient, the model demonstrated robustness to dynamic environments and the simulations were more realistic.

## 4.2 Simulation Models

Simulation models combine different well-known techniques and methods focusing on the development of novel general behavior models, without creating a new system or tool. Those models are usually implemented as a plug-in or extension of an existing simulation system, making it more realistic and efficient, or adding new features to it.

In 2009, VEROSIM system[13] was exploited for the development of a prototype model for generic multi-agent-based simulations [86]. The agent's behavior was based on the human behavior representation, the steering behavior (changes in direction of an agent) was based on some of those proposed by Reynolds' [87] (described by Buckland [88]), the locomotion used forward Euler integration and the pathfinding used Dijkstra's pathfinding algorithm [89]. The results showed that the agents' behaviors were realistic, the simulation software was efficient and the model could simulate 2000 agents in real-time.

In 2018, Kasereka et al. proposed a model for the design and simulation of evacuation of people from a building on fire [90], which was based on four parameters (total people alive, total deaths, average potency of alive people and total time taken) focusing on a practical evaluation of the evacuation process. The model was developed over the GAMA Platform. Although many factors were not taken into account (like emotion, stress, age etc.), the model could be used on several types of commercial buildings without major changes. On the other hand, Karbovskii et al. presented a multi-model agent-based simulation technique that incorporated multiple modules [91], consisting of informational planning, decision-making mechanisms, navigation and collision avoidance methods. They also proposed a method of model integration using common virtual space and abstract common agents. The simulations were done using the PULSE simulation tool [92]. From the simulations' aspect, it was highlighted that the time step increased linearly when a single model was used, but changed when the number of models increased. Also, there was an overhead in network data transfer and it required additional empirical research to calibrate it correctly.

Liu et al. presented a simulation approach that uses the multi-population algorithm framework, combining the Cultural Algorithm [93] and a proposed improved Social Force Model (SFM) [94] [95]. The proposed approach divided the population space into groups and selected a leader for each one. The proposed SFM had advantages over the original in cases with obstacles and multiple exits, but the simulations were conducted with a relatively simple scene. Lastly, in 2020, Badeig et al. proposed a new model, called Environment as Active Support for Simulation (EASS), that improved the configurability of the simulation process by delegating the context computation process to the scheduling process [96]. The EASS was implemented as a plugin for the Madkit MAS platform [97] and the results showed that the EASS was computationally more efficient, more flexible and simpler.

[10] An annual Islamic pilgrimage to Mecca, Saudi Arabia. (https://en.wikipedia.org/wiki/Hajj)
[11] A multimethod simulation modeling tool developed by The AnyLogic Company (https://www.anylogic.com)
[12] A toolkit for developing and comparing reinforcement learning algorithms. (https://gym.openai.com)
[13] A 3D simulation system for environment, industry and space simulation. (https://www.verosim-solutions.com)

## 4.3 Methods and Techniques

In this section we analyze the methods and techniques exploited in CSCM by presenting a brief outline for each one and by mentioning any technologies used for its implementation or simulation, along with its advantages and drawbacks.

### *4.3.1 Navigation*

The most important and resource consuming part of the CSCM is the navigation of the agents in a microscopic level. Navigation takes into account the whole environment and has to determine how the agent will move or which path to take to reach a specific goal. These methods and techniques are focused on the collision avoidance, navigation behavior (changes in navigation due to narrow passages or high density), trajectory calculation and path finding.

Starting with the collision avoidance methodologies that have been proposed, in 2009 Karamouzas et al. presented a local method for collision avoidance based on collision prediction [4]. This method was relatively easy to implement, yielded real-time performance and characters showed smooth behavior and avoid each other in a natural way. Despite that, this method required a manual calibration to work appropriately. The same year, Guy J.Stephen et al. presented a local collision avoidance algorithm between multiple agents, called ClearPath [98]. The algorithm was implemented using the RVO-Library [99], over the OpenSteer [87] multi-agent simulation system. The authors extended the RVO's formulation by imposing additional constraints, with Eq. 1 and 2 showing the two boundary constraints, taking into account a truncated cone (FVO).

$$FVO_{LB}^{A}(v) = \varphi\left(v, \frac{v_A + v_B}{2}, p_{ABleft}\right) \geq 0 \quad (1)$$

$$FVO_{RB}^{A}(v) = \varphi\left(v, \frac{v_A + v_B}{2}, p_{ABright}\right) \geq 0 \quad (2)$$

Where $A$ and $B$ are two objects, $p_{ABleft}$ and $p_{ABright}$ the inwards directed rays perpendicular to the left and right edges of the cone respectively, $v_x$ is the velocity of the object and $\varphi$ the distance between two points. The algorithm was performing well on different kinds of scenarios, complex or not, with a very high number of agents. Despite that, there were cases where there might have be a collision free path, but the algorithm was not be able to compute it, also the computation of a new velocity for each agent could change agent's behavior of the algorithms for the rest of the simulation.

An approach for vision-based collision avoidance between walkers that fit the requirements of interactive crowd simulation was presented in 2010, [100]. The algorithms were implemented using the OpenGL [101] and CUDA libraries, simplifying the geometries of the environment and simulating four different examples to show its improvements. The results showed that there was an improvement in the emergence of self-organized patterns of walkers, with future promising results. On the other hand, computational complexity of the proposed algorithms was too high, highlighting it as fairly complex systems and difficult to be customized for different situations/environments. Golas et al. proposed an algorithm for extending existing collision avoidance algorithms to perform better approximation and long-range collision avoidance, while also proposing a metric for quantifying the smoothness of agents' trajectories [102]. The extension was applied using the RVO2 Library [103], removing its default restriction of maximum neighbors. With the proposed algorithm, crowds reached their goals faster with speeds similar to real people, which presented more realistic behaviors, but also cases of outlying pedestrian behaviors could be observed appropriately, such as a pedestrian having different speed compared to the others. On the other hand, in 2016 Li and Wong presented an agent-based approach for 2D crowd simulation that, using the agent's radial view, could compute a set of collision free movement directions [104]. This method was simple, fast and performed well in environments with simple obstacles. An end-to-end framework for reactive collision avoidance policy generation for efficient distributed multiagent navigation was introduced by Long et al., [105]. They used the Optimal Reciprocal Collision Avoidance (ORCA) algorithm [51] from the RVO2 library to generate data for the training of a Deep Neural Network (DNN), supplying it with frames showing how an agent should avoid its surrounding agents. Although, the DNN did not completely fit the data, the learning policy generalized well to unseen situations and was easier to be exploited than the ORCA algorithm. It should be highlighted that the authors did not use any static obstacles to increase the complexity of the environment.

Moving to the navigation behaviors, in 2009, Narain et al. presented an approach for crowd simulation that used dual representation, the first one (discrete) being a single agent and the second one (continuous) being a crowd [1]. For the latter, they introduced a variational constraint (unilateral incompressibility) that modeled large-scale collision avoidance and accelerated inter-agent collision avoidance in scenarios where the crowd was dense. The proposed method could handle multi-agent simulations with crowds of hundreds of thousands of agents efficiently, but could not avoid collisions with distant agents. Later, in 2014, Best et al. presented an algorithm to model density-dependent behaviors, called DenseSense [106], that aimed to generate pedestrian trajectories using the Fundamental Diagram of traffic flow[14] [12]. The simulations were carried out using the Menge crowd-simulation framework. The proposed approach had a small computational overhead, generated smoother trajectories, was applicable to a large number of global (*algorithms that compute the path towards the goal*) and local (*algorithms that modify the trajectory of a path to avoid collisions*) multi-agent algorithms and generated realistic density-dependent human-like behaviors. On the other hand, its benefit may be diminished in scenarios where global density-dependent navigation techniques were already used. The same year, Bastidas proposed a different kind of simulation method that used RL algorithms for low-level decisions during navigation [107]. The algorithms focused on training the agent to reach a goal-point while avoiding obstacles that might be on its way while trying to follow a coherent path. After the agent was trained, the knowledge was transferred by using the resulting Q-Table (*a lookup table, which stores the maximum future reward of each action and is used to choose the best action for each state*) [108]. The method fulfilled their goals to a certain degree but failed to represent certain situations, for example it did not take into account situations where agents could not move, so in some cases agents moved over other agents or obstacles. In 2017, Dutra et al. proposed a perception/motion loop to steering agents along collision free trajectories while also introducing a cost function based on perceptual variables that estimated an agent's situation considering both the risks of the collision and a desired destination [109]. The proposed method showed realistic human-like behaviors. The developed agents' behaviors were commonly observed in the day-to-day life and agents kept more natural distance between them. On the other hand, an appropriate parameter setting was required for different scenarios and the behaviors tested took place on environments with static obstacles. In the same year, a method for the detailed modeling of agent behaviors in Lagrangian formulation [110] was introduced by Weiss et al. [111]. They modified the Position-Based Dynamics scheme [112] and added additional constraints for short and long-range collision avoidance. The implementation of the framework was done in CUDA library, running at interactive rates for hundreds of thousands of agents.

Path planning and trajectory calculation are some of the hottest points in CSCM. In 2010 Guy et al. focused on trajectory calculation [113] with promising outcomes. They presented an optimization algorithm that computed paths based on the Principle of Least Effort and generated energy-efficient trajectories based on biomechanical principles, for agents in crowd simulations. The simulations were implemented in C++ and used OpenGL libraries for the visualization. The presented algorithm could simulate thousands of agents and automatically generated many emerging different behaviors. Despite that, the measurements they did, were based only on humans walking in straight lines at normal speeds. Humans were represented by hard disks of fixed radius (ignoring the fact that sometimes people may “squeeze”) and the implementation of the algorithms could not compute the optimal trajectory. In 2011, Viswanathan and Lees proposed two additions to the RVO library: a) Group sensing for motion planning, to avoid groups of agents, and b). Filtering of percepts, based on interestingness, to model the limited information processing capabilities of human beings [114]. The results showed that the proposed collision avoidance algorithm created more realistic group avoidance behavior. In 2016, Support Vector Machines (SVMs) combined with SFM were used to produce destination paths faster in dense crowds [115]. The same year, Bera et al. used a genetic parameter learning algorithm for data-driven crowd simulation and content generation, that was based on incrementally learning pedestrian motion models and behaviors from crowd videos [116]. They integrated an improved tracking algorithm based on Particle Filters [117], with an RVO based motion model, improving accuracy and producing smoother trajectories. Despite that, the algorithm did not model aspects such as physiological and psychological traits or age and gender and it did not work well in some more complex cases. On the other hand, in 2017, Wong et al. presented an algorithm to compute the optimal route for each local region for evacuation simulations, by reducing the congestion and maximizing the number of evacuees arriving at exits in each time span [118]. The proposed algorithm could handle crowds with various attributes in various environments that were difficult for previous methods. Also, the algorithm was flexible and adaptable to elevators and public transportation. The same year, Han et al. focused on agent routing but with a different approach [119]. They proposed an extended Route Choice Model (RCM) to simulate the way pedestrians selected an appropriate route, among the Available Evacuation Route Set (AERS), during an evacuation process. The RCM, AERS and a Modified SFM (MSFM) were combined to avoid obstacles and to select an appropriate route. In this context they used a RL method to optimize the routes in the AERS. The results showed that it could reproduce realistic crowd dynamics and route choice behaviors. By setting its parameters, the relationship between congestion and scenario and decomposition of congestion required data mining. Lastly, in 2019, an improved multi-agent RL algorithm was introduced to solve the problem of mutual influence of agents in path planning-based crowd simulation [120]. Additionally, they improved the original SFM by adding a cohesive force of visual factors to its mathematical formula (Eq 3). The proposed algorithm could effectively improve the evacuation efficiency.

[14] A diagram that gives a relation between the traffic flux (vehicles/hour) and the traffic density (vehicles/km)

$$m_i \frac{\overrightarrow{d v_i(t)}}{dt} = \overrightarrow{f_i^0} + \sum_{j(\neq i)} \overrightarrow{f_{ij}} + \sum_{w} \overrightarrow{f_{iw}} + \sum_{j(\neq i)} \overrightarrow{f_{ij}^{rel}} \quad (3)$$

Where $\overrightarrow{v_i}$ is the actual speed of movement of the pedestrian, $\overrightarrow{f_i^0}$ the self-driving force of $i$ driven by the direction of the target, $\overrightarrow{f_{ij}}$ the force between an individual and another individual, $\overrightarrow{f_{iw}}$ is the force between an individual and an obstacle and $\overrightarrow{f_{ij}^{rel}}$ the force of attraction between members of a group.

The application of well-known methodologies that focus on the navigation in general (methodologies/algorithms etc. applied in other similar problems) have shown important flourishment the last decade. In 2013, Dutta and McLeod used a very simple Least Effort Model [121], along with a cell based environment, to simulate and study large crowds and their interactions with each other and the environment [122]. They used MATLAB[15] and CUDA[16] library for the simulations, simulating up to about 160.000 agents, but lacking realism. A year later, a pedestrian simulation framework (MARL-Ped) was introduced [123], that used a model-free RL algorithm (*RL algorithms that do not use the transition probability distribution and reward function associated with the Markov Decision Process (MDP)* [67]) for virtual environment navigation. Each agent used this model, which took raw environment information, like the agent's velocity, angle and distance by employing tile coding [67], [124], for value function approximation. The agents were independent and did not communicate with each other, with the model controlling their velocity, driving them towards their goal. The results showed that the proposed method could solve different problems (problems such as shortest vs quickest path, two groups crossing a corridor and pedestrians walking through a maze). Also, it was highlighted that the algorithm learned pedestrian behaviors, like detouring of peripheral agents due to high density or creation of congestion in corridors. On the other hand, the result of the model could not be modified, making it an issue for behavioral animations. Continuing their work, they combined the vector quantization with Q-Learning and different iterative learning strategies [6]. In this approach agents learned independently. The proposed framework was applicable and generalizable and the created behaviors seemed to be scalable. In 2015, Boatright et al. presented an approach for agent steering that uses machine-learned policies [125]. The policies were based on the behavior of a procedural steering algorithm, through the decomposition of the possible steering scenarios. Firstly, they generated the required data for the training using an oracle algorithm [126], which included the agent density and net flow near the agent. Then they used multilevel Decision Trees (DTs) [127] for the training of the policies. The proposed approaches showed massive increase in efficiency with higher population numbers and had low number of collisions, but the DTs aspect was too restrictive in the case of chosen action was incorrect. In 2018, an agent-based deep reinforcement learning approach was introduced, where only a reward function enabled agents to navigate in various complex scenarios/environments, with a single unified policy for every scenario [128]. For the model's input they used a set of consecutive depth maps measured by the agents. Despite that, the agents could not be distinguished from obstacles and in some cases the resulting trajectories were not those of the shortest path. In 2019, Hildreth and Guy proposed an algorithm for coordinated multi-agent navigation (cooperative multi-agent systems) by training agents to use decentralized communication [129]. The authors applied the MAS algorithms along with Time-to-Collision Forces (TTC-Forces) algorithms for collision avoidance and Particle Swarm Optimization (PSO) approaches [130] for the learning process. The results showed that the agents learned meaningful communication that improved their performance in dynamic environments (non-stationary environments), but they were not so effective in the case of transferring the model to scenarios (environments). In addition to that, PSO had limitations when applied to this kind of issues (knowledge transferring). On the other hand, an effective data-driven crowd simulation method was presented, that could mimic the observed traffic of pedestrians in a given environment by using the observed trajectories [5]. The authors used Generative Adversarial Networks (GANs) [131] to learn the properties and generate new trajectories with similar features, while also trying to effectively handle local collision avoidance. They combined GANs with flexible route following algorithms that take into account temporal information and, as a result, the system could be used in real-time. Also, they mentioned that the introduced methods could be combined with other simulation methods by allowing more realistic interactive applications. On the other hand, the proposed method could generate trajectories without taking into account other agents. Similarly, using NNs, Oshita proposed a method for individual agent control using a Deep Convolutional Neural Network (D-CNN) [132]. The proposed model learned from a heat map that contained the positions and speeds of nearby agents and the temporary target position that indicated an appropriate heading direction, so that the agent reached the final position efficiently.

[15] A multi-paradigm numerical computing environment and proprietary programming language developed by MathWorks. (https://www.mathworks.com/products/matlab.html)
[16] A parallel computing platform and API model created by Nvidia. (https://developer.nvidia.com/cuda-downloads)

#### *4.3.2 Agent behavior*

Agent behavior analysis are very important and complicated studies of CSCM in microscopic level. In CSCM studies, two types of behaviors are key reference points, individual behaviors and group behaviors. Individual behaviors are focused on agent behaviors that are influenced by their traits, preferences or other decision process methods, while group behaviors are focused on the behavior of a crowd during movement, like retaining their formation or group.

Decision process is one very challenging and difficult task. Luo et al. presented the HumDPM, a decision process method for agents, that incorporated experience and emotion [133]. Thus, the decisions were made by matching past experience cases to the current situation. The authors used stages, feature cues, set of goals, experiences and actions for the design of experience. The emotion elicitation process they used was based on the Appraisal Theory [134]. Khouj et al. showed some prelaminar results of the modelling and simulation of an intelligent agent by using RL algorithms to assist a human emergency responder, with a goal of maximizing the number of patients discharged from hospitals or on-site emergency units [135]. During the simulations, the proposed agent, called DAARTS (Decision Assistant Agent in Real Time Simulation), was able to help the emergency responder, leading to a favorable outcome. In 2015, Fu et al. proposed a model, for crowd evacuation, that integrated multi-agent technology into CA, by also combining it with a perceptual and decision model that they designed [136]. The perception model took into account visual information. The decision-making process included different behaviors during the movement towards the target location, like herding (Eq. 4), local collision avoidance, escape behavior, random behavior (moving randomly), helping behaviors and many others. The results were very encouraging and showed that the model simulated crowd behaviors realistically and effectively.

$$\begin{cases} \vec{v}_{prefer} = \lambda(\alpha\vec{v}_{separate} + \beta\vec{v}_{align} + \gamma\vec{v}_{cohesion} \\ \vec{v}_{separate} = \left[\sum_{i=1}^{n} \frac{1}{d(i,A)}\vec{v}_i\right]\eta_1 + \vec{v}_A(1-\eta_1) \\ \vec{v}_{align} = \left[\sum_{i=1}^{n} \frac{1}{n}\vec{v}_i\right]\eta_2 + \vec{v}_A(1-\eta_2) \\ \vec{v}_{cohesion} = \left[\sum_{i=1}^{n} \frac{P_i}{n} - P_A\right]\eta_3 + \vec{v}_A(1-\eta_3) \end{cases} \tag{4}$$

Where $\vec{v}_{separate}$, $\vec{v}_{align}$ and $\vec{v}_{cohesion}$ are the velocities generated by separate, alignment and cohesion behavioral rules, $\eta_1$, $\eta_2$ and $\eta_3$ are control parameters, $\lambda$ is the agent's HP and $\alpha, \beta$ and $\gamma$ are set according to the confusion in the environment.

By taking advantage of the characteristics of crowds in indoor environments, Pax and Pavón presented a flexible agent decision model for indoor environments to address requirements such as scalability and performance issues and flexibility of an agent model [137]. Each agent had its own high-level behaviors (reactive plans) and low-level behaviors (speed, position, angles, density). The high-level behaviors were responsible for *what do to next* and their design followed the Behavior Oriented Design (BOD) method [138], while the low-level behaviors carried out the operations. The simulations were conducted using the MASSIS framework, implementing efficient algorithms for path finding, localization of elements (QuadTree [139]) and steering behaviors.

Popular RL methods, such as Sarsa-On Policy Temporal Difference, have been exploited for the development of a decision-making algorithm for agents in emergency response scenarios [140]. Lopez et al. conducted the simulations using the I2Sim simulation tool [141] and the results' performance was quite higher compared to their previous studies. Liao et al. proposed a methodology for multi-agent simulation systems [142], that exploited the Bayesian-Nash Equilibrium [143] for the decision-making process. The method was calibrated and validated using data collected from real experiments. It could help design optimal walkway width (safety-wise and flow rate-wise) and pedestrian flow rate, but did not take into account pedestrian psychology, social relationships, information transmission and obstacles during the evacuation.

In general, crowd (or group) behaviors, represent more realistically the actions of humans in dangerous situations. Sun and Wu focused on crowd's behavior heterogeneity, presenting a method that simulated the individual's behavior that increased the heterogeneity in the crowd simulation for more realistic results [144]. The model was based on two physics models, Reynolds' Boids and Helbing's Social Force [94], [145]. Also, they introduced a number of parameters that could be used to configure the crowd behavior. Having the same aim, Guy et al. presented a technique to generate heterogeneous crowd behaviors using personality traits theory and proposed a novel two-dimensional factorization of perceived personality in crowds [146]. The simulations were done by using the RVO2 library. From the outcomes, a mapping of the relationship between crowd simulation parameters and perceived personality traits was derived. This approach could successfully generate simulations where agents had various levels of the established personality traits and could perceive more than 95% of the captured data. On the other hand, the implementation explored only variations that were allowed by the library and focused mainly on local behaviors and interactions. In 2012, Liu et al. presented a novel algorithm [147] that utilized the Discrete Choice Model (DCM) [148], combining it with the Dynamic Feedback Model (DFM). By exploiting the capabilities of the DCM and the DFM, they implemented goal selection in crowd behavior for situations like evacuation, shopping and rioting. In the case of an evacuation, Eq. 5 shows the probability of an agent choosing exit $i$ while Eq. 6 shows the attractiveness of the exit,

$$P_{i,a} = \frac{e^{\frac{\beta_0}{d_{i,a}}+\beta_1 A_i}}{\sum_j e^{\frac{\beta_0}{d_{j,a}}+\beta_1 A_j}} \tag{5}$$

$$A_i = v_{pos} e^{k_{pos}(n_i/n_{i,max})} - v_{neg} e^{k_{neg}(1-\sqrt{\frac{n_{i,max}}{n_i}})} \tag{6}$$

where $\beta_0$ is the weight of the distance between the agent and the exit, $\beta_1$ the weight of the attractiveness of an exit, $n_i$ the number of agents near the exit, $n_{i,max}$ the number of agents that could block the exit, $v_{pos}$ a constant to scale the expression result within $[0, 1]$, $k_{pos}$ a constant to calibrate positive feedback and $v_{neg}$ and $k_{neg}$ are constants related to the degree of influence of the negative feedback. This algorithm could simulate heterogeneous crowd behaviors and worked well with local collision avoidance models (such as RVO). On one hand, the system was limited to certain crowd scenes but on the other hand it could provide control of variables.

Yan et al., focused on the believability/realism of the crowd's behavior and simulated and investigated it via three multi-agent RL methods [8]. The first method adopted the Q-Learning with Multi-agent MDP (MMDP) and the other two methods, that were introduced, exploited the joint state action Q-learning approaches and the joint state value iteration algorithm. The simulations were conducted in a simple grid world environment. All methods showed promising results and produced believable behaviors. Agents in MMDP were experiencing collisions presented some process difficulties, for this reason the number of agents had to be greatly reduced. Peymanfard and Mozayani proposed a holonification method (joining of an agent into a crowd) for data-driven crowd modelling [149]. Using real-world data and Random Forests (RFs) they modeled the rules of joining each agent to a crowd and leaving it. The simulation results confirmed the generalization of the proposed method was very fast, could be used in a real-time simulation and could provide more realistic experimental outcomes. Heliövaara et al. presented a model that tried to represent human-like behaviors in counterflow situations where they try to avoid collisions with oncoming agents [150]. They implemented the system in the FDS + Evac system (Fire Dynamic Simulation) [151]. In their model, agents were able to dodge multiple agents at a time in a realistic way by improving the original model and prevent unrealistic jams. Focusing on the behavior derived from the agent communication system, Kullu et al introduced an approach that simulated human-like communication methods between agents [152], that corresponded to a simplified version of the Foundation for Intelligent

Physical Agents[17] (FIPA) and the Agent Communication Language[18] (ACL) message structure specifications [153]. The simulations were implemented and generated using the Unity game engine and showed that the proposed approach improved the behavioral variety of grouping behaviors in emergency situations, while simulation trajectories were more realistic. However, despite the fact that agents traveled less, it took more time to evacuate and the flow rates from a passageway scenario showed that the communication could not cause significant change in such cases. On the other hand, SOLACE, a multi-agent method for human behavior during seismic crisis [154], incorporated geographic information and social bonds, that was based on the social attachment theory, with Eq. 7 showing how the perception distance for a bond is calculated.

$$PD_{Bond} = PD_{Normal} * (1 + \frac{1}{10} * SD_{Bond}) \tag{7}$$

where $PD_{Normal}$ is the agent's normal perception distance and $SD_{Bond}$ the social distance. This algorithm was adaptable and during the simulations, using the perception aided by social bonds, the agent movements appeared realistic. Despite those, there was a need to focus on improving its realism.

### *4.3.3 Emotional aspects*

Realism is another important aspect of the CSCM systems and emotion integration is a very important feature of those simulations. The emotional aspect of the systems focuses on simulating emotion contagion and different kind of emotion models that influence the agents' behavior.

One of the first emotion models that balanced physiology and cognition to create realistic characters was introduced in 2010 [155]. For the creation of the agent's behavior and to model their mental states, the authors implemented the BDI paradigm, the i* method [156], the JADEX framework [157] and jMonkey[19] for the graphics. In addition to that, they applied a RL method for the navigation of the agents. Although, the proposed approach did not allow the modelling of all basic human characteristics, it seemed to be very fast. Two years later, Li et al presented a method for agent distribution using psychological preferences [158]. This method exploited the Centroidal Voronoi Tessellation approach [159], in combination with the Truncated-Newton algorithm to construct the method, and the OCEAN model for the development of the agents' personality traits. They combined the OCEAN model with Maslow's Hierarchy of Needs approach [160] for the needs and environment features, by tagging areas with additional information (functionality, size and density). The simulations showed realistic and efficient crowd distributions, providing agents with the ability to choose preferred goal locations. Each agent chose an area by using the functionality (Eq. 8), size (Eq. 9) and density (Eq. 10) equations:

$$S_{func} = \frac{\sum_{k=1}^{5} \left\{ weight_k \times \frac{\psi_i^O + \psi_i^C + \psi_i^E + \psi_i^A + \psi_i^N}{0.2} \right\}}{0.2} \tag{8}$$

$$S_{size} = \omega \times \left\{ (\psi_i^E + \psi_i^N) \times 0.5 \right\} \tag{9}$$

---

[17] FIPA is an IEEE Computer Society standards organization that promotes agent-based technology. http://www.fipa.org/

[18] ACL is a proposed standard language for agent communication. https://en.wikipedia.org/wiki/Agent_Communications_Language

[19] A game engine for 3D development. (https://jmonkeyengine.org)

$$S_{den} = \delta \times \{(\psi_i^O + \psi_i^C + \psi_i^E + \psi_i^A + \psi_i^N) \times 0.25\} \quad (10)$$

$$S_{overall} = \alpha_{func} \times S_{func} + \alpha_{size} \times S_{size} + a_{den} \times S_{den} \quad (11)$$

Where $\psi$ is the personality factor, $\omega$ is the region size scale (1 to 5), $a_{func}, a_{size}, a_{den} \in (0,\ 1)$ are the weights of each equation with $a_{func} + a_{size} + a_{den} = 1$. The region with the highest $S_{overall}$ was chosen by the agent.

The last years, emotion contagion showed high flourishment and in 2017 Ta et al. proposed an emotion contagion method based on social psychology [2], which was implemented in the GAMA, multi-agent-based simulation platform. The authors assessed the impact of emotion decay, of the environment and of the agents' neighbors on the dynamics of the emotion, at individual and group levels. Another emotion contagion method was presented by Cao et al. [3], which combined the OCEAN and the Susceptible Infected Susceptible (SIS) models [161] to develop a novel P-SIS (Personalized SIS) model. The different variables used in the model were defined in Eq. 12-16. Their experimental results showed realistic pedestrian flows, with future promising results.

$$d_i = log - N(\mu_d, \sigma_d^2) \quad (12)$$

$$T_i = log - N(\mu_T, \sigma_T^2) \quad (13)$$

$$D(i, R) = \sum_{j \in Neighbor(i)} d_j \quad (14)$$

$$E_i(0) = P_N(i) E_i(0) \quad (15)$$

$$E(t) = E(t-1) - \beta E(t-1) \quad (16)$$

For Eq. 12 and Eq. 13, where $d$ is the panic dose, $T$ the threshold of accumulated panic dose, $i$ the individual, $\mu_d$ mean of $d$, $\sigma_d^2$ variance of $d$, $\mu_T$ mean of $T$, $\sigma_T^2$ variance of $T$ and $d$ and $T$ are calculated by a pre-determined log-normal distribution function $log - N$. For Eq. 14, $R$ is the distance threshold for neighbors, $i$ and $j$ the individuals and $D(i, R)$ is the accumulated panic dose of individual $i$ from its neighbors $Neighbor(i)$. For Eq. 15 and Eq. 16, $\beta$ is the emotional decay factor, $t$ the time step, $N$ the number of individuals in the crowd and $E(t)$ the emotion value of infected individuals at time $t$. The emotion value corresponds to an emotional state, with 0 being calm, $(0, 0.4]$ anxiety, $(0.4, 0.8]$ panic and $(0.8, 1.0]$ hysteria.

A targeted study on specialized human-like stress feature was presented by Rockenbach et al. [162]. It was introduced as a method to parametrize crowd simulation allowing the increase or decrease of agents' stress. Moreover, they proposed an extension for the BioCrowds [163] system to deal with agents' comfort and stress..

#### *4.3.4 Group dynamics*

Large-scale simulations are also part of the CSCM literature and simulate large crowds of agents, usually at a macroscopic level. Those methods and techniques focus on the group's navigation, collision avoidance, formation maintenance and adaptation to available space.

In 2010, Qiu and Hu introduced a methodology based on utility and social comparison theories [164]. They incorporated a two-layer framework, where the top layer described the context of group formation and the bottom the step of individual selection. The results showed that it could simulate dynamic grouping and that the social factor had great impact on crowd behaviors. Continuing their work [165], they presented a framework for modeling the structure aspect of different groups of pedestrian crowds, intra-group structure and inter-group relationship. The system was implemented in an agent-based crowd behavior simulation over the OpenSteer environment. The individual groups' structure could be dynamically changed based on the agents' spatial distance, agents' similar goal and agents' social proximity. As a consequence, there was a need of performance improvement for large-scale group-based crowd simulations. Focusing on another aspect of group dynamics, Ju et al. presented a method that blended existing crowd data to generate new crowd animations [166]. The proposed method, learned crowd movements from observed data and could then generate spatially larger crowds that look perceptually similar to the input data. The crowd model could include an arbitrary number of agents for an arbitrary duration, leading to a natural looking mixture of crowds that could have predictable control over various crowd styles/types. Some structures, though, could not be captured faithfully by the system, such as a case of a herd of gnus with a sparse formation, which was captured correctly but the temporal variation of the formation could not be represented. Additionally, the model did not take environment features into account (except collision avoidance) and large crowds were cumbersome for the system. By taking into account scalability and performance, Wang et al clustered agents to crowd groups towards the development of a more effective crowd simulation algorithm [167]. The simulations' results showed that the presented algorithm was effective when obvious flow patterns were shown within the crowd and could also cluster agents according to their long-term interests. In 2016, Nasir et al introduced a technique for the simulation of large groups in dense crowds [168], focusing on the change of the groups' formation for avoiding collisions and maintaining the agents' collective behavior. They used a leader-followed methodology, a modified SFM, to maintain the group's formation (type) and agent density. Eq. 17 shows how a slot of a formation could be chosen by a follower.

$$p_{i,j} = p'_{leader} - iS_r v + jS_C(v \times z) \tag{17}$$

Where $p_{i,j}$ is the $(i, j)$ follower (of a total $m$ and $n$ respectively), $v$ the orientation vector, $z$ the unit upward vertical vector, $p'_{leader}$ the future leader and $S_r$ and $S_c$ are the standard intervals for rows and columns of the group respectively. The method can be applied to individual agents to simulate group formation changes depending on the density of the agents. It should be highhearted that in some cases the group might choose incorrect pathways to pass through.

BioCrowds algorithm was presented to study crowd's behaviors by A. D. L. Bicho et al. [169]. The algorithm was based on a space colonization algorithm to model leaf venation patterns and branching in trees. The algorithm applied the competition for space, collision avoidance and lane formation to the crowd, taking into account relationship of crowd density and speed of agents. The proposed algorithm generated guaranteed collision-free motion based on markers that could be used to interact with the crowd (follow specific paths or change the crowd's density). The environment was monitored in a simple way where each agent could observe free space rather than observing other agents. In contrast to BioCrowds, in 2015 Jaklin et al. presented a method for the simulation of the walking behavior of small pedestrian groups [170], called Social Groups and Navigation. They defined the group coherence and sociality of a crowd by using a vision-based collision avoidance method [171] with some modifications and a SFM [172]. From one hand, the simulation results showed that the method's performance was almost the same for different group sizes and could be easily parallelized. From the other hand, avoidance behavior was not so effective in respect to entire groups. Simply put, it did not give control over the splitting behavior and the agent's vision was not influenced by the environment.

A decentralized algorithm for group-based coherent and reciprocal group navigation, generating macroscopic group movements, was introduced in 2016 [173]. The algorithm included agent clustering, inter-group and intra-group proxemic avoidance. The algorithm had a number of benefits and resulted in smooth and coherent navigation behaviors of groups, but had a 10-14% of runtime overhead. The same year, Zhong et al. proposed a data-driven modeling framework to construct agent-based crowd models using real-world data and to predict the trajectories of the pedestrians (agents' groups) [174]. Quite similarly to [164], Zhong et al. exploited a dual-layer architecture in which the bottom layer included collision avoidance behaviors and the top layer included goal selection and path navigation methods. The results showed that the framework generated behaviors effectively and offered promising performance on future trajectory predictions. Despite those results, the velocity initialization in complex scenarios needed improvement and the knowledge learned from a video was suitable only for a specific environment. In 2019, Ruiz and Hernández introduced a GPU-based hybrid method for crowd simulation that exploited RL algorithms to guide groups of pedestrians towards a goal by adapting the groups' dynamic to environmental dynamics [175]. Additionally, they presented a CA method to describe the interactions of each pedestrian. The simulations were visualized using an engine implemented in C/C++ and OpenGL. The method supported different behaviors through MDP layers and the approach allowed the setting of dynamic goals and obstacles to which the crowd adapts during simulation time.

Regarding the group formation and organization, in 2011 an agent-based methodology for the explicit representation of groups of pedestrians that form crystals of crowds[20] was presented, [176]. The agent's behavioral rules were derived from the Proxemics theory [177] with some changes to the metrics. In 2018, Zhang et al. proposed a modified two-layer SFM for grouping and guidance, using crowd dynamics and extended the study of group organization patterns to a higher density [178]. The modified SFM included group partitioning, leader selection, modeling for leaders, group members and disorganized pedestrians that join a group. The movement of disorganized pedestrians is described with Eq. 18.

$$m_i \frac{d\vec{v_i}(t)}{dt} = (1-\beta_i)\vec{f_i^0} + \beta_i \sum_{g\in G_i} \xi(g)\vec{f_{ig}^0} + \sum_{j\neq i}\vec{f_{ij}} + \sum_{w}\vec{f_{iw}} \tag{18}$$

$$\xi(g) = \frac{count[N_{ig}]}{\sum_{g\in G_i} count[N_{ig}]} \tag{19}$$

$$\beta_i = \begin{cases} 0, exit_i \in \Omega_i \\ \dfrac{\sum_{g\in G_i} count[N_{ig}]}{count[N_i]}, exit_i \notin \Omega_i \end{cases} \tag{20}$$

Where $\vec{f_{ij}}$ illustrates the interaction forces from the pedestrian $i$ and pedestrian $j$, $\vec{f_{iw}}$ the interaction force between pedestrians and walls, $\Omega_i$ the range of disorganized pedestrian $\iota$'s vision, $exit_i$ is the exit that $i$ chooses, $G_i$ is the aggregation of groups within the range of vision, $N_i$ the aggregation of pedestrians within the range of vision, $N_{ig}$ the aggregation of group $g$ within the range of vision and $\vec{f_{ig}^0}$ refers to the members of the closest group. The modified model had more stable group organization and fewer shocks and was more efficient in realizing better evacuation. On the other hand, Collins and Frydenlund investigated strategic group formation through the introduction of cooperative game theory techniques into an agent-based model and simulation by using a Core-inspired approach for exploring strategic group information [179]. This model gave the benefit of empirical results with the limitation of standard cooperative game theory and opened a new research field in CSCM

*4.3.5 Other*

[20] Small, rigid groups of men, strictly delimited and of great constancy, which serve to precipitate crowds.

In addition to the above categories, the CSCM literature has focused on other, smaller, parts of the simulations that are as important as the above, such as performance improvements or parameter tuning.

A Genetic-Fuzzy System for agent steering, that automated the tuning process of the rules was introduced in 2010, [180]. The simulation results showed that such setup was very useful and efficient for reducing the human interface with the simulation environment for further tuning. Also, it was highlighted that the algorithms could operate dynamically in a real time simulation. In 2011, Navarro et al. introduced a framework for dynamic level of detail for large scale simulations, that could be applied to all agent models [181]. Their aim was to simulate areas of high level of interest more precisely using microscopic models and other areas less precisely but more efficiently using macroscopic models. The framework included a dynamic change of representation, focusing on the navigation between one model to another dynamically. Additionally, a spatial aggregation method was applied to represent agents at a less detailed level, by considering similar agents as a group of individuals with similar psychological profiles and common physical space. The implementation was done in SE-*, a Thales proprietary multi-agent simulation system [182]. The proposed approach showed lower CPU resources usage when simulating a high number of agents and could automatically determine the most suitable representation level for each agent. On the other hand, the framework used fixed parameters leading to small aggregation size, the scalability was over-simplified and in cases with high density of agents the aggregation was less precise. In 2012, S. Lemercier et al. presented a numerical model that simulated following behaviors that are present in situations like queueing and walk-in corridors [183]. They calibrated their model through an experimental approach, which consisted of capturing the motions of people performing following and stop-and-go movements, with different number of participants and length of corridor. The calibrated model, which was inspired by the Aw-Rascle model [184], could fit a range of densities and was more suited to control the speed when there were constraints.

A novel algorithm for the development of a model for physics-based interactions in multi-agent simulation environments was introduced in 2013, [185]. The algorithm was capable of modeling both physical forces, interactions between agents and obstacles, allowing agents to anticipate and avoid collisions for local navigation. The proposed method provided stable, anticipatory motion for agents while incorporating agent responses to forces and could easily be combined with other approaches. Assuming that agents were constrained to moving along a 2D plane using a simple 2D circle for the approximation of each agent, it was highlighted that the method could not be physically accurate. Focusing on Data-Driven Agent-based Modelling, in 2019, Malleson et al. incorporated a Sequential Importance Resampling Particle Filter with systematic resampling [186] for real-time data assimilation [187]. Their aim was to show the reliability of the use of particle filters to estimate the "true" state of a system, using an agent-based model and observational data. For the simulations, they used the StationSim model[21] and the results showed that, it was possible to perform dynamic adjustments of the agent features but the increase of the number of agents required an exponential increase of the number of particles in the Particle Filter.

# 5 Discussion

In this section, the aforementioned methodologies, models and systems are compared in a way to provide the reader a comprehensive and comparative review of the literature, as well as to establish a general framework for future studies by highlighting some of most impactful key elements of CSCM. It should be noted that for each category the works are compared based on similar features. For example, in the category of Navigation (Section 4.3.1), all works can be compared according to the number of the agents they can simulate and the speed of the simulation. On the other hand, methodologies that have to do with the simulation of the agent's behavior (Section 4.3.2) or the emotional aspects (Section 4.3.3) of the agents, could not be compared based on the same features, population size and computational requirements.

[21] Repository containing the source code to run the StationSim model. (https://github.com/urban-analytics/dust)

Table 1 presents an analysis of the works related to crowd simulation tools. The comparison criteria are the following:

1. Key feature: some of the most important key features of the simulation system.
2. Simulation size: the number of agents the system can handle.
3. Parallelized: if the system has implemented parallel computing for the simulation to increase the experimental speed.
4. Platform (system): the exploited platform for the development of the simulations system. We use the "custom" term in case the authors developed their own system for the simulations.

Table 1 Comparison of crisis simulation systems and tools.

| Reference | Key feature | Simulation Size | Parallelized | Platform |
|---|---|---|---|---|
| [48] | Movement of characters (agents) with vastly different sizes and ability to move under each other (when possible). | Not defined | Yes | Custom |
| [50] | Simulation of large scale tawaf with agents of different physical capacity and activity | 35.000 | Yes | Custom |
| [52] | Perception and interpretation of signage information | 1.700 | No | Unity |
| [53] | Real time simulation of 1 million agents | 1.000.000 | Yes | Custom |
| [55] | Utilization of Hadoop cluster | 10.000 | Yes | Custom |
| [58] | Functionality is highly extensible, independent elements co-exist, scalable with agent size and its behavioral finite state machine | 64.000 | Yes | Menge |
| [59] | Ability to specify different crowd types | 100 | No | Unity |
| [63] | NVRAM incorporation for high number of agent simulation | 100.000 | Yes | Custom |
| [65] | NVRAM incorporation for high number of agent simulation | 60.000 | Yes | Custom |
| [30] | GAML language, BDI agents and can simulate various types of scenarios | * | Yes | GAMA |

* Not defined

From the above table, it seems that the focus of the crowd simulation systems tends to differ from one to another. Although three of the presented systems focused solely on simulating as many pedestrians as possible, all the others focused on different things. [58], though, also focused on the functionality of the system and its usability, making it highly extensible and easy to use, which is rarely considered when developing such a system. In addition to that, it should be highlighted that most simulation systems develop their own tool, and some of them make use of popular game engines, such as the Unity. On the other hand, the simulated number of the agents vary based on the needs of the case study.

A short comparison of the crisis simulation tools is presented in Table 2. The comparison criteria are the following:

1. Key feature: some of the most important key feature of the simulation tools.
2. Simulation scenarios: the scenario the system can simulate.
3. Simulation size: the number of agents the system can handle.
4. Parallelized: if the system has implemented parallel computing for the simulation.
5. Platform (system): the exploited platform for the development of the simulations system. We use the “custom” term in case the authors developed their own system for the simulations.

Table 2 Comparison of crisis simulation systems and tools.

| **Reference** | **Key feature** | **Simulation scenarios** | **Simulation Size** | **Parallelized** | **Platform** |
|---|---|---|---|---|---|
| [68] | Use of temporal-difference learning for the simulation of heterogenous agents | Large building evacuation with spreading fire | 2 | No | MATLAB |
| [69] | Combines rule reasoning with numerical calculation | Fire evacuation | * | No | Custom |
| [70] | Use of temporal networks for city evacuation hit by tsunami | City evacuation hit by tsunami | * | No | MATSim |
| [72] | Includes evacuees and resources and entities contributing to the evacuation | Three story building evacuation | 30 | Yes | Custom |
| [16] | Different agent types, visualization for planning and training and spread of realistic knowledge. | Airport evacuation | 200 | * | ESCAPES |
| [74] | Realism of environment | Building evacuation | * | * | Unity |
| [75] | Identifies best locations for the exits and most effective position of signs | Fire emergencies | * | No | EvaPlanner |
| [76] | Support for security services and use uses optimized techniques | Arena evacuation | 22.500 | Yes | Evakuierungsassistent |
| [78] | Use of CA that represent different levels of environment | Soccer stadium evacuation | 21.874 | No | Custom |
| [79] | Concert venue scene | Fire disaster and evacuation | * | No | Custom |
| [82] | Use of ANNs | Evacuation | * | No | Custom |
| [7] | Environment created in OpenAI gym and uses new training approach | Fire evacuation | * | No | Custom |
| [85] | RL based, data-driven simulation, cohesiveness-based K-means for trajectory grouping and merging | Evacuation | * | No | Custom |

* Not defined

Looking at the above table, it is obvious that in most cases, there is no clear mention of the achievable simulation size. Additionally, most of the simulation scenarios tend to be evacuations, or evacuations with a spreading fire, which makes the simulation more complex. Once again, we can see that custom-made simulation systems are mostly used, followed by the use of game engines. An important issue that should be highlighted is that, while parallelized computing approaches are very important in CSCM studies, they are rarely adopted. This may be due to the difficulty of implementing such algorithms/methods.

The following criteria focus on the comparison of the simulation models and the results are analyzed in Table 3, criteria:

1. Key feature: some of the most important key features of the simulation tools.
2. Simulation scenarios: the scenario the system can simulate.
3. Simulation size: the number of agents the system can handle.
4. Platform (system): the exploited platform for the development of the simulations system. We use the "custom" term in case the authors developed their own system for the simulations.

Table 3 Comparison of simulation models.

| **Reference** | **Key feature** | **Simulation scenario** | **Simulation Size** | **System** |
|---|---|---|---|---|
| [86] | Flexible modeling of agent's cognitive behavior and usage of optimization techniques for the simulation of high number of agents | Densely-packed environments, theatre filling, fire evacuation, cityscape and ancient infantry battle | 2.000 | VEROSIM |
| [90] | Intelligent agents and parameters for evacuation evaluation | Supermarket fire evacuation | * | GAMMA |
| [95] | Two-layer control mechanism (belief and population), knowledge based evacuation, group leaders and improved SFM | Teaching building evacuation | 300 | MonoGame |
| [91] | Crowd pressure metrics, common space for the interaction of different agent | Cinema building evacuation | 400 | PULSE |
| | models and commonly controlled agents | | | |
| [96] | Optimization of context computation and configuratble design process and simple reusability of behaviors in different simulations | Transportation crisis | * | Madkit |

* Not defined

The results of the Table 3, show that the evacuation in different cases is once again the focal point of presented of the works. Specifically, the evacuation of large indoor facilities.

The analysis of the agent navigation approaches follows in Table 4. The comparison criteria chosen for navigation approaches are:

1. Simulation size: maximum number of agents that could be simulated.
2. Simulation update speed: the simulation speed of the methodology.
3. Simulation scenario: the scenario simulated in which the proposed methodology was used.
4. Simulator: simulation system or any other framework used for the simulations.
5. Parallelized: if the methodology takes advantage of parallel computation.
6. ML Model: machine learning approaches that was used.

Table 4 Comparison of proposed methodologies and techniques for navigation.

| Reference | Simulation Size | Simulation Speed | Simulation scenario | Simulator | Parallelized | ML Model |
|---|---|---|---|---|---|---|
| [4] | 1.000 | 30 FPS | Park | * | Yes | No |
| [98] | 250.000 | Real time | City | Custom | Yes | No |
| [1] | 100.000 | 2.23 FPS | Campsite | * | No | No |
| [100] | 1.000 | 4.3 FPS | Walkers examples | Custom | Yes | No |
| [113] | 10.000 | 58.9 FPS | Long Corridor | Custom | Yes | No |
| [122] | 100.000 | * | Crowds change side | MATLAB | Yes | No |
| [102] | 2.000 | 200 FPS | 4-way crossing | * | No | No |
| [125] | 3.000 | 10 FPS | Urban | * | No | DT |
| [104] | 1.600 | * | Blocks | * | No | No |
| [115] | 400 | * | Two ways corridor | * | * | SVM |
| [116] | * | * | * | Custom | No | Bayesia n |
| [109] | 300 | 10 FPS | Crossing | Custom | Yes | No |
| [111] | * | 50 FPS | Crossing | * | Yes | No |
| [118] | 35.000 | * | Exhibition Evacuation | * | No | No |

* Not defined

The outcomes of the Table 4 seem to show that the collision avoidance algorithm Guy J.Stephen et al. [98] presented is the best choice, as it can simulate the highest number of agents (250.000) in real time and also make use of parallelized approaches for simulating large city blocks. Beside the works presented in Table 4, there are many other that are analyzed in Section 4.3.1 but not included, due to their differentiation based on the chosen criteria.

The works presenting agent behavior methodologies are compared in Table 5, where some details of regarding approaches are analyzed. Those details are:

1. Focus: what behavior the technique focused on simulating.
2. Simulation scenario: the scenario simulated in which the proposed methodology was used.
3. Simulator: simulation system or any other framework used for the simulations.
4. Parallelized: if the methodology takes advantage of parallel computation.
5. ML Model: if the proposed methodology uses a machine or deep learning model.

Table 5 Comparison of proposed methodologies and techniques for agent behaviors.

| Reference | Focus | Simulation scenario | Simulator | Parallelized | ML Model |
|---|---|---|---|---|---|
| [133] | Behavior based on experience and emotion | Subway station evacuation, food distribution | 3D game engine | No | No |
| [135] | Train agent for decision making | Production cells monitoring | i2Sim | No | RL |
| [136] | Perceptual and decision models for evacuation | Evacuation | * | Yes | No |
| [140] | Coordinated decision making | Venue evacuation | i2Sim | Yes | RL |
| [144] | Increase heterogeneity of crowd | Exit from a corridor | * | No | No |
| [146] | Heterogeneity of crowd behavior using personality trait. | Passthrough, hallway, narrowing passage | * | No | No |
| [147] | Goal selection and heterogeneity of crowd behavior | Evacuation, tradeshow tour, rioting | RVO library | No | No |
| [8] | Believability of crowd behavior | Grid world | Custom | No | RL |
| [137] | Flexible decision model for indoor scenarios | University evacuation | MASSIS | No | No |
| [149] | Crowd holonification | * | Custom | No | RF |
| [150] | Behavior model for counterflow situations | Counterflow in corridors, intersection, merging flows | FDS + Evac | No | No |
| [152] | Communication between agents | Bidirectional flow, passageway, evacuation, chat | Unity | No | No |
| [154] | Behavior based on social attachment theory | City evacuation | GAMMA | No | No |

* Not defined

Most of the simulation scenarios are navigation focused, meaning that they simulate scenarios where the agents or crowds are moving to a specific destination, where the decision process and behavior plays a higher role in the outcome. Additionally, most of those use an existing platform, applying the behavioral methodologies to an existing algorithm or platform. Contrary to the methodologies and techniques presented for the navigation, there are quite a few that make use of a machine learning model. Naturally, most of them made use of RL, except of one that used RF. Additionally, only two of the methodologies had a part of them parallelized, with [136] having only the movement parallelized, while [188] being fully parallelized.

The presented methodologies regarding the emotional aspect of the simulations present various results, so it is difficult to categorize them as the previous approaches. This makes sense, because this field is quite active, new and requires specialized skill. Some of those focused on behavior change due to an emotional state and others on emotion contagion, with the main focus of the literature being the emotion contagion. Some highlights of the methodologies are:

- The implementation of the BDI paradigm by Da Costa et al. resulted in agents with realistic behaviors and improvised actions, having real time performance [155].
- W. Li's et al. showed that by implementing goal preferences into the agents, the crowd distributions were more realistic [158].
- Very realistic results regarding the crowd density in cases of panicked or calm people were shown by introducing a stress level model (better than the model used in BioCrowds) [162].
- In the general view, it seems that the use of ML models is not an option when the modeling of emotion is the goal

The research sub-domain of group dynamics in the simulations is also a very new area with few works. For this reason, we highlight only some key points. Those, focused on the group formation, navigation, behavior and other aspects, presenting different results for each case. From the presented literature we can derive the following:

- ML models are not used often for the modeling of group dynamics, despite the fact that there are methodologies focusing on different aspects of crowd dynamics (navigation, behavior, formation, dynamics).
- Computational parallelization methodologies are not exploited often due to implementation difficulties. Specifically only two of the presented approaches were parallelized, Jaklin's et al. walking behavior [170] and Ruiz & Hernández's GPU-based implementation of the model [175].
- Simulation size does not seem to be a big concern in crowd dynamics modeling, so the number of the agents vary based on the case. For example, D. L. Bicho's et al. presented that BioCrowds algorithm could simulate 800 agents at 30FPS and 13.000 at 6FPSin a test case of interactive crowd control [169], while Jaklin's et al. simulated 1980 agents at 20FPS in a room evacuation scenario.
- Algorithms such as K-means seem to be very effective in large-scale evacuation scenarios [167].
- Group navigation showed high performance in a holonomic and two non-holonomic simulations with small numbers of collisions [173]

The study of the "other" aspects (subfield of the Methods and Techniques category) presented several various novel research contributions. For this reason, we present some of the most important key elements:

- Behavior level of detail studies seem to perform ever well  in subway station and large city scenarios [181], with 100-300 agents. On the other hand, when the number of the agents increases to 500-1000 and 10.000 the efficiency of the methods decreases, especially in the case of the large city.

- A new introduced custom crowd simulation model [183], based on the Aw Rascle model, showed very promising results in the domain of the distributions of pedestrians with stop-and-go waves, compared to older models such as Helbing [145], RVO2 [145] and Tangent [189]
- Physics-based interactions can simulate high number of agents at high frame rates when integrated with the Bullet Physics Library [185].
- The use of a particle system showed that it can incorporate external data into the system dynamically and the mean error variance between the true data and the estimated data is lower when the number of particles increases [187].

# 6 A General Framework

Taking into account the above presented works and the highlights that emerged from the comparative study, we introduce a general framework for the future CSCM studies/developments focusing in high level outcomes with huge social and scientific impact. For this reason, we propose a list of key elements, which will enrich this framework and will be considered as coordination/guide mechanism for further studies in field of multi-agent based CSCM. The following key factors will be considered:

- 3D simulation platform: Crowd and crisis simulation systems tend to focus on game engines for the development of the system. This is done due to the fact that they offer a wide variety of tools that are directly correlated with the simulations, such as physics simulation, 3D environment building etc. There are numerous game engines available, but Unity has been used most.
- 2D simulation platform: This type of simulations should be considered only in cases where the scenarios are required to be studies in a higher level of interactions and movements. Although 2D simulations are less hardware intensive, simulations (crowd or crisis) in general require high level of realism and should simulate physics as realistically as possible. As 2D simulations are simpler than 3D, a custom-made system can always be considered and is considered in many cases. In other cases, Unity already offers tools for the creation of 2D environments.
- Agent navigation: As seen in the current review, the literature has focused greatly on the navigation. It is one of the most important parts of a simulation system and coupled with obstacle avoidance and other behaviors, it can be a very performance intensive procedure. The algorithm that will be used must be efficient and should be able to perform in real-time, as simulations tend to have high numbers of agents moving and performing actions. Due to the fact that there is a need to simulate as many agents as possible, working on the efficiency of those kind of algorithms will be done in the future.
- Group management: In all cases of crisis situations, humans tend to form groups and act together and for this reason, group management algorithms are also an important part of a CSCM system.
- Agent behavior: The agent's behavior is another important part and should model microscopic decisions made by the agent. The behavior, although in a microscopic level, can add an important detail of realism which will eventually make the system's simulations more reliable. Although ML models can yield very efficient results, with the appropriate training process, it should be taken into account that humans do not always act efficiently, even more in crisis situations. Emotion aspects: the emotional aspect of the simulations makes them more realistic and useful for crisis scenarios and a number of different methodologies should be applied. Emotion contagion is a known and studied part of the emotional aspect, adding important behavioral changes. Moreover, the literature has shown that by incorporating the OCEAN and SIS models, crisis simulations become more realistic.

- Population size and performance: When designing a simulation scenario, the population size is an important factor. It heavily depends on the system's hardware performance, the exploited algorithms and the scenario itself. Despite that, literature shows that the system should be able to handle indoor scenarios with about 1.000 agents and 3.000 agents in outdoor scenarios. Higher number of agents would be preferable without limitations, to represent realistically huge experiments.
- Machine learning approaches: The continuing research has proved that the application of machine learning is an important part of many AI applications. For this reason, the incorporation of ML models in different parts of the simulation of crowd for crisis management will yield better results. An important ML paradigm that has been applied and shown in the presented analysis is the RL. Two widely used RL libraries are OpenAI and Stable Baselines[22]. Additionally, as Unity is used quite often as a simulation platform, the ML-Agents[23] toolkit should be considered.
- Simulation parallelization: The use of parallelization techniques for real time simulation systems is an important part. All standard parallelization techniques and tools make use of the CPUs and are provided with most of the development frameworks, but Open MPI[24] is a widely used library for this purpose. The use of the GPU, though, can increase the performance of the simulations much more and CUDA is a widely used library for exploiting Nvidia GPUs processing power. In many cases ML libraries, such as scikit-learn [190] support CUDA. For this reason, ML libraries as well as simulation tools supported by CUDA should be preferred.
- Modification support: It is highlighted the need for novel systems that are able to support the ability of third-party modifications. Modifications are alterations done by other users to a system that aim to change one or more aspects of it. This part has become especially important the last years as there are many studies that aim to extend the functionalities of other systems, incorporate new features to existing systems or improve existing ones.
- Dynamic environment development/management: A very important feature of a novel simulation system is the creation and management of dynamic environments. Naturally, not all scenarios (real or not) have the same environment and environmental structure. This enables the user to simulate a plethora of different scenarios, with different building architectures and environmental designs etc. This approach will give users a powerful tool, an easy adaptable environment to any crisis case.
- Dynamic crisis management: Another important part towards a novel and more efficient simulation system should include the dynamic change of crisis and its behaviors. For example, after an earthquake may follow a tsunami, or after a fire a flood may cause severe issues. A system providing the modeling of such crisis pipelines would be effective in simulation approaches. This kind of functionalities, also may include the ability of dynamically adding new issues to the crisis scenario, for example starting a fire in a different part (than the initial) of the environment. In addition to that, this functionality can be used to test newly implemented agent behaviors.

[22] A set of improved implementations of reinforcement learning based on OpenAI Baselines. https://github.com/hill-a/stable-baselines
[23] Open-source Unity plugin that gives the ability to train intelligent agents. https://github.com/Unity-Technologies/ml-agents
[24] Open source MPI implementation. https://www.open-mpi.org/

# 7 Conclusions

In this paper, we presented the trends of the last decade in the field of crowd simulation for crisis management. With the growth of technology and computation power, the interest in simulations has grown and one of the hottest topics of simulations has become crowd and crisis simulations. The first one is to simulate the crowd, along with its behaviors, psychology and movements, in both indoor and outdoor aspects like festival, trading centers or stations. The latter is to simulate the same things, but for crisis cases, like flooding, fire or earthquake evacuation scenarios, due to the fact that these cases are extremely complex and have huge social impact.

The studied literature in this paper showed that this topic has been an issue for several decades, with many methodologies being proposed through the years, with a slow start at the beginning due to the limitation of computational resources. As the computation power increased in the past decade, the crowd simulation topic has been getting more and more interest and the need for novel simulation tools, methods, algorithms and methodologies flourished.

By providing an extensive analysis of the methodologies, techniques, models, systems and tools that have been proposed in the last decade, we presented them in a categorical and in a comparative way. This revealed the most effective aspects in crowd simulation for crisis management via multi-agent systems. These aspects are provided as key elements for future studies which can be considered as a general framework to be followed by scientists. It is clear that as years pass, and the computational resources increases:

1. multi-agent-oriented aspects are considered to be very effective,
2. the number of simulated agents increases exponentially,
3. game engines are adopted for more realistic outcomes,
4. complex machine learning algorithms provide better human-like representations, especially RL algorithms, which are considered to be the best reflection in human behaviors,
5. huge indoor and outdoor environments are simulated easily and
6. low-level human behaviors studies are considered to be very important.

By exploiting the study of this paper, one can choose the appropriate methodologies and techniques that fit better the requirements of the simulation system they intend to build. Also, it is evident that the field still has to find its own forum in the scientific publication realm. So far published literature is greatly scattered in different journals, conferences and workshops; the field could benefit from special issues in established scientific journals and dedicated workshops in multi-agent systems conferences.

To sum-up, this study can be considered as a key reference point for future research, development and demonstration of multi-agent-based crowd simulation applications in the crisis management domain.

# 8 Acknowledgments

This work is partially supported by the MPhil program "Advanced Technologies in Informatics and Computers", hosted by the Department of Computer Science, International Hellenic University.